\documentclass{article}

\usepackage{PRIMEarxiv}

\usepackage{subfig}

\usepackage[utf8]{inputenc} % allow utf-8 input
\usepackage[T1]{fontenc}    % use 8-bit T1 fonts
\usepackage{hyperref}       % hyperlinks
\usepackage{url}            % simple URL typesetting
\usepackage{booktabs}       % professional-quality tables
\usepackage{amsfonts}       % blackboard math symbols
\usepackage{nicefrac}       % compact symbols for 1/2, etc.
\usepackage{microtype}      % microtypography
\usepackage{lipsum}
\usepackage{fancyhdr}       % header

\usepackage{graphicx}       % graphics
\graphicspath{{media/}}     % organize your images and other figures under media/ folder
\usepackage{amsmath}
\usepackage{xspace}

\let\OldTexttrademark\texttrademark
\renewcommand{\texttrademark}{\OldTexttrademark\xspace}%
\title{An enhanced Conv-TasNet model for speech separation using a speaker distance-based loss function
}

\author{
  Jose A. Arango-Sánchez \\
  Intelligent Information Systems Lab - In2Lab \\
  Universidad de Antioquia \\
  Medellín, Colombia.\\
  \texttt{jose.arangos@udea.edu.co} \\
   \And
  Julián D. Arias-Londoño \\
  GAPS-SSR \\
  ETSIT-Universidad Politécnica de Madrid \\
  Madrid, Spain.\\
  \texttt{julian.arias@upm.es} \\
}

\begin{document}
\maketitle

\begin{abstract}
This work addresses the problem of speech separation in the Spanish Language using pre-trained deep learning models. As with many speech processing tasks, large databases in other languages different from English are scarce. Therefore this work explores different training strategies using the Conv-TasNet model as a benchmark. A scale-invariant signal distortion ratio (SI-SDR) metric value of 9.9 dB was achieved for the best training strategy. Then, experimentally, we identified an inverse relationship between the speakers' similarity and the model's performance, so an improved ConvTasNet architecture was proposed. The enhanced Conv-TasNet model uses pre-trained speech embeddings to add a between-speakers cosine similarity term in the cost function, yielding an SI-SDR of 10.6 dB. Lastly, final experiments regarding real-time deployment show some drawbacks in the speakers' channel synchronization due to the need to process small speech segments where only one of the speakers appears.
\end{abstract}

% keywords can be removed

\section{Introduction}
Speech is a natural form of interaction between humans and systems (computers, cell phones, voice assistants, etc.). However, the speech of interest is often obscured by various natural environment sounds, such as background noise, other interfering speakers, the reverberation of the environment, etc., which hinder the interaction between humans and systems. For cases in which the audio to be processed corresponds to the mixture of different speakers, speech separation systems have been proposed as valuable alternatives to isolating the individual interventions of the speakers \cite{opensourceseparation:book}. For humans and especially for the auditory system, speaker separation is a job of great ease; however, it has proven to be very difficult to build an automatic system that equals or surpasses the human auditory system at this task \cite{wang2018supervised}.

With a wide range of applications in diverse fields such as medicine with hearing aids, commercial hearing aid design, mobile telecommunications, and automatic and robust speech and speaker recognition \cite{wang2018supervised}, speaker separation is a field of interest and continuous development. However, as in the vast majority of voice and speech processing tasks, most of the existing approaches have been developed and trained with datasets of the English language and under controlled conditions regarding noise types and levels. Therefore, in this work, we explore the performance and behavior of a well-known model architecture \cite{luo2019conv} designed for speaker separation, evaluating it with a database of call recordings over telephone channel in the Spanish language with speakers from different parts of Latin America. This scenery poses a challenge because of the database's rich diversity of Spanish accents. Besides, the telephone channel recordings are limited to a 3KHz bandwidth which degrades signal quality. 

Furthermore, after evaluating the model's performance, an inverse relationship between the quality of the separation and the speakers' similarity was found, which led to the introduction of a term in the cost function related to lowering the similarity between the estimated sources, in a similar fashion as Domain Adversarial Training \cite{ganin2016domain} reduces the model's bias regarding a context variable. The speaker similarity was estimated using pre-trained speech embeddings, allowing for improve the performance of the speaker separation task in terms of a scale-invariant signal distortion ratio.

The rest of the paper is organized as follows: section \ref{sec:state-of-the-art} introduces the speech separation field in more detail and describe the most relevant works published in recent years. Section \ref{sec:methods} presents a description of the baseline architecture and the proposed modification. After, section \ref{sec:exper} describes the experimental setup and the results obtained for the different variants evaluated. Lastly, section \ref{sec:coclusions} presents some conclusions derived from the results.

%Finally, the modified Conv-TasNet model was deployed in an infrastructure that allowed its execution in real time.

\section{Related work}\label{sec:state-of-the-art}

Traditionally, speaker separation has been tackled from signal processing, however, during the last decade, with the introduction of deep learning, this task has been approached as a learning problem, where speech, speaker and noise discrimination patterns are learned during training, thus emerging approaches (algorithms, models and architectures) that have drastically accelerated the progress and performance of supervised speaker separation \cite{wang2018supervised}. These approaches have allowed the development of a wide range of applications in diverse fields such as medicine, hearing aids, mobile telecommunication, automatic and robust speech and speaker recognition. among others \cite{wang2018supervised}.

Current approaches for speaker separation based on deep neural networks are trained with a large number of audios, composed of a mixture and the different isolated sources. Since it is necessary to have the mixture and the sources, many of the speaker separation systems are supervised machine learning systems. The neural network must produce an output for each source (estimated source) and then compare each output with the actual isolated source during the training process. This comparison is used as a loss function to update the network's weights inside the back-propagation algorithm.

Depending on the number of sensors or microphones, speaker separation methods are classified as monaural (single-microphone) and array-based (multi-microphone) separation methods. During this work, we will deal with monaural methods due to the nature of the corpus to be used (telephone call recordings). Most of the methods developed have formulated the separation problem through a time-frequency (T-F) representation or spectrogram of the mixed signal, which is estimated from the waveform using the short-time Fourier transform (STFT). Therefore, these methods aim to estimate a clean spectrogram of the individual sources from the spectrogram of the mixture. This process can be performed using two approaches \cite{luo2019conv}:

\begin{itemize}
\item Direct method. It performs the direct estimation of the spectrogram representation of each source in the mixture using non-linear regression techniques, where the spectrograms of the clean sources are used as training targets.
\item Mask estimation method. It estimates a weighting function (mask) for each of the sources that, when multiplied by each Time-Frequency (T-F) bin of the mixture spectrogram, recovers the individual clean sources.
\end{itemize}

Originally, both approaches calculate the waveform of each source using the inverse short-time Fourier transform (iSTFT) from the estimated magnitude spectrogram of each source, along with the original phase. The mask estimation approach is the most common in recent years and in which deep learning has achieved the most significant increases in terms of separation metrics. However, this approach has several shortcomings, the most notable of which are \cite{luo2019conv}:

\begin{enumerate}
\item The STFT is a generic signal transformation, which is not necessarily optimal for speech separation.

\item The time-frequency representation requires a high-resolution frequency decomposition of the mixture signal, which requires a long time window to ensure successful separation. Therefore, it increases latency and computational cost, limiting its use in real-time applications.

\item The dissociation between magnitude and phase creates a challenge at the time of performing waveform reconstruction.
\end{enumerate}

Given these difficulties, a fully convolutional time-domain audio separation network (Conv-TasNet) was proposed in \cite{luo2019conv}, which follows an End-to-End approach, and uses a linear encoder to generate an optimized speech waveform representation, to later be used in the separation of individual speakers. The Conv-TasNet linear encoder replaces the STFT, which is used for feature extraction, by a data-driven representation optimized as part of an End-to-End training paradigm. The final speaker separation is achieved by applying a set of weighting functions (masks) to the encoder output, and the modified encoder representations are then converted into waveforms by a linear decoder, which plays the role of the iSTFT.

The Conv-TasNet model generates the masks using a temporal convolutional network (TCN). This network is made up of stacked 1-D dilated convolutional blocks, which allows it to model long-term dependencies of the speech signal while maintaining a small model size. This model was evaluated using objective distortion metrics such as scale-invariant signal-to-noise ratio (SI-SNRi) and signal-to-distortion ratio (SDRi). Also, through subjective measures of audio quality such as perceptual evaluation of subjective quality (PESQ) and mean opinion score (MOS).

The Conv-TasNet model used the WSJ0-2mix (two speakers) and WSJ03mix (3 speakers) datasets for training. These datasets are commonly used by most of the state-of-the-art speaker separation approaches. They were generated by mixing randomly selected English audios from different speakers extracted from the Wall Stree Journal (WSJ0) \cite{Hershey_2016} dataset, which were then mixed with random noise between -5 dB and 5dB. The performance of the Conv-TasNet model with the WSJ0-2mix dataset was 15.3 dB for SI-SNRi, 15.6 dB at SDRi, MOS of 4.03, and a PESQ of 3.22. Finally, having a significantly small size (5.1 Million parameters) makes it a suitable solution for both offline and real-time speech separation applications. However, since Conv-TasNet uses a fixed temporal context length, long-term tracking of an individual speaker may fail, especially when long-duration pauses are present in the audio to be separated \cite{luo2019conv}.

Usually, time-domain-based speaker separation systems receive input sequences with a large number of time steps, which introduces challenges when modeling extremely long sequences. An evident alternative is to use recurrent layers instead of convolutional ones, but difficulties would arise in the network's optimization. On the other hand, if one-dimensional convolutional networks (1-D CNNs) are used, modeling of sequences at the utterance level would not be possible when the receptive field is smaller than the sequence length \cite{luo2020dual}. Therefore, to solve these difficulties, a double-path recurrent neural network (DPRNN) was proposed in \cite{luo2020dual}, which calls itself a simple method that organizes the layers of an RNN into a deep structure that allows extremely long sequences to be modeled.

The DPRNN approach divides the input sequence into shorter chunks and interleaves two RNNs, an intra-chunk RNN and an inter-chunk RNN, which perform local and global modeling, respectively. In a DPRNN block, the intra-chunk RNN processes the local chunks independently, and then the inter-chunk RNN aggregates information from all chunks, thus achieving utterance-level processing. Finally, this model uses the same database as the previous approach (WSJ0-2mix), obtaining 18.8 dB in the SI-SNRi metric and 19.0 dB in SDRi \cite{luo2020dual}. However, RNNs are inherently sequential models; therefore, parallelizing their calculations is impossible, and they tend to generate a bottleneck when processing large data sets with long sequences. Considering recent advances in deep learning, a natural alternative to standard RNNs is replacing recurrent calculations with multi-head mechanisms. Thus was born SepFormer, a neural network based on transformers, which learns short- and long-term dependencies, with a multi-scale approach to speaker separation \cite{subakan2021attention}.

The SepFormer network, like the two previous approaches, works under the time domain, adopting the framework proposed in DPRNN. It replaces the recurrent components with a multi-scale pipeline composed of transformers. This approach achieves the highest performance among the two previous approaches; using the same dataset (WSJ0-2mix) it obtained 22.3 dB in SI-SNRi. 

%Finally, all the approaches found in the state of the art address the problem of speaker separation as a supervised learning problem, using single-microphone separation methods. Each of the proposals found, tries to solve a deficiency present in the same state of the art: Conv-TasNet replaces the STFT, which is used for feature extraction, by a data-driven representation, the DPRNN introduces the RNN since the Conv-TasNet (CNN 1-D) does not perform a modeling of sequences at utterance level, on the other hand, the SepFormer when using transformers, allows the parallelization of calculations and avoids bottlenecks generated by the RNN.

Notwithstanding the results of recent approaches, we use Conv-TasNet as the benchmark architecture to evaluate a Spanish language speech separation task in this work.  Our interest in the Conv-TasNet architecture is backed for the following reasons: 1) It significantly outperforms masking methods based on time-frequency representations for two- and three-speaker mixtures and assessed by both objective measures of distortion and subjective quality assessment; 2) Conv-TasNet has a significantly small model size (5.1 Million parameters) and minimal latency compared to other models, making it a suitable solution for real-time speech separation applications \cite{luo2019conv}, and also for experiments with limited hardware.

\section{Methods}\label{sec:methods}

\subsection{The Conv-TasNet model}

\begin{figure}[!h]
    \centering
    \includegraphics[scale=0.38]{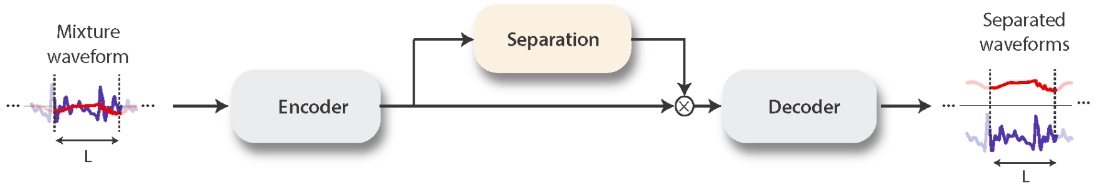}
    \caption{Key components of the Conv-TasNet architecture. Taken from \cite{luo2019conv}.}
    \label{fig:tasnet_block}
\end{figure}

The Conv-Tasnet architecture is composed of 3 key components, which are shown in Figure \ref{fig:tasnet_block}. Globally, this architecture has an Encoder, which generates a high-dimensional representation of segments of the mixture waveform. Then, we have a Separation component, which computes a mask for each target source. Finally, a Decoder reconstructs the waveforms of the sources based on the masked features \cite{luo2019conv}.

\begin{enumerate}
    \item The Encoder transforms short segments of the mixture waveform into an intermediate feature space using a 1-D convolution operation, thus replacing the STFT for feature extraction with a data-driven representation, which is optimized together with an End-to-End training paradigm.
    
    \item The Separation module estimates a multiplicative function (masks) for each source at each time step. These masks are obtained through a temporal convolutional network (TCN), which is formed by 1-D stacked dilated convolutional blocks. This allows the network to model the long-term dependencies of the speech signal while maintaining a small model size. Finally, speaker separation is achieved by applying a set of weighting functions (masks) to the encoder output.

    \item The Decoder reconstructs the waveform using a 1-D transposed convolution operation, thus replacing the iSTFT.
    
\end{enumerate}

\begin{figure}[!h]
    \centering
    \includegraphics[scale=0.5]{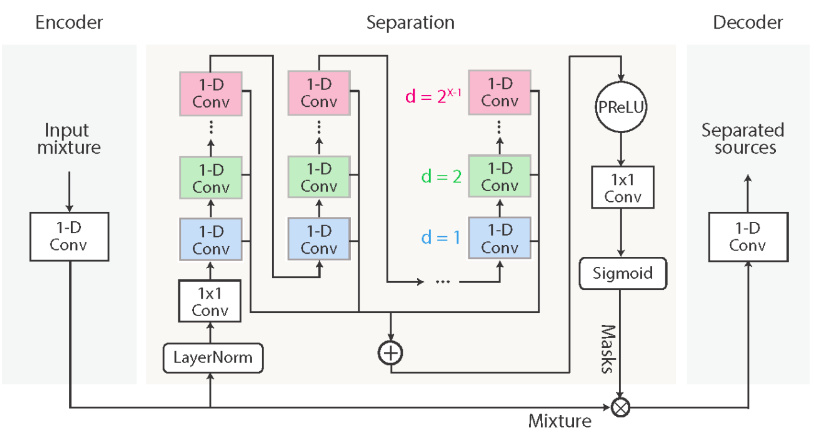}
    \caption{Conv-TasNet architecture flowchart. Taken from \cite{luo2019conv}.}
    \label{fig:system_flowchart}
\end{figure}

\begin{figure}[!h]
    \centering
    \includegraphics[scale=0.6]{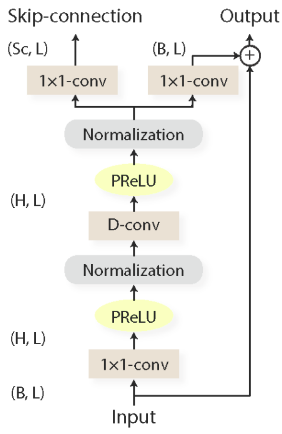}
    \caption{Structure of a convolutional block (1-D Conv) ConvTasNet architecture. Taken from \cite{luo2019conv}.}
    \label{fig:block_conv}
\end{figure}

As shown in Figure \ref{fig:system_flowchart}, both the Encoder and the Decoder are composed of a 1-D convolution block, which is described in Figure \ref{fig:block_conv}. Where each block consists of a 1×1 convolution operation followed by a depth-wise convolution operation, with a nonlinear activation function and a normalization added between each of the two convolution operations. Finally, two 1 × 1-conv linear blocks serve as residual path and jump connection path for the next block, respectively \cite{luo2019conv}.

For its part, the separation module, which can be seen in Figure \ref{fig:system_flowchart}, is in charge of estimating the masks based on the encoder output. This module is composed of different layers, which in turn contain different 1-D convolutional blocks. Each 1-D convolutional block has different dilation factors, which increase exponentially. This guarantees a sufficiently large temporal context window, thus taking advantage of the long-range dependencies of the speech signal. The different colors of the 1-D convolutional blocks denote different dilation factors \cite{luo2019conv}.

The cost function used by the Conv-TasNet model corresponds to the maximization of the scale-invariant Signal-to-Noise Ratio (SI-SNR). The SI-SNR is defined by:

\begin{equation}
    \text{ SI-SNR : }=10 \log _{10} \frac{\left\|\mathbf{s}_{\text {target}}\right\|^{2}}{\left\|\mathbf{e}_{\text {noise }}\right\|^{2}}
\end{equation}

where $\mathbf{s}_{\text {target }}$ and $\mathbf{e}_{\text {noise }}$,  are given by:

\begin{equation}
    \mathbf{s}_{\text {target }}:=\frac{\langle\hat{\mathbf{s}}, \mathbf{s}\rangle \mathbf{s}}{\|\mathbf{s}\|^{2}}
\end{equation}

\begin{equation}
    \mathbf{e}_{\text {noise }}:=\hat{\mathbf{s}}-\mathbf{s}_{\text {target }}
\end{equation}
$\hat{\mathbf{s}} \in \mathbb{R}^{1 \times T}$ and $\mathbf{s} \in \mathbb{R}^{1 \times T}$ are the estimated source and true source, respectively, and
$\|\mathbf{s}\|^{2}=\langle\mathbf{s}\mathbf{s}\rangle$ denotes the signal power. Scale-Invariant is guaranteed by normalizing $\hat{s}$ and $s$ to zero mean before calculation \cite{luo2019conv}.

Finally, the most commonly used evaluation metric for speaker separation systems is the Scale Invariant Signal Distortion Ratio (SI-SDR), which is considered a measure of how good an estimated source sounds, in other words, it measures the ``overall quality'' of the estimated sources.

Before defining such a metric, we must mathematically define what makes up an estimated source: 

Let $\hat{s}_i$, be the estimated source of the $i$-th speaker, which is composed of four elements\cite{opensourceseparation:book}:

\begin{equation}
    \hat{s}_i = s_{\text{target}} + e_{\text{interf}} + e_{\text{noise}} + e_{\text{artif}}
\end{equation}

Where $s_{\text{target}}$ is the true source, $e_{\text{interf}}$, $e_{\text{noise}}$ and $e_{\text{artif}}$ are the error of the added interference, noise and artifacts, respectively \cite{opensourceseparation:book}. Detailed calculation of these terms can be found in \cite{vincent2006performance}.

Once these terms are defined, the signal distortion ratio is calculated as follows:

\begin{equation}
    \label{eqn:sdr}
    \text{SDR} := 10 \log_{10} \left( \frac{\| s_{\text{target}} \|^2}{ \| e_{\text{interf}} + e_{\text{noise}} + e_{\text{artif}} \|^2} \right)
\end{equation}

To achieve scale invariance, a rescaling process is performed prior to the calculations.

\subsection{Similitude between speakers}

After an initial set of testing experiments using a pre-trained Conv-TasNet model provided by its authors, and with different audios from the corpus of calls (see section \ref{sec:exper}), it was perceptually identified that, for speakers with similar voices, all the former models obtained poor results compared to when the voices of the involved speakers sounded different. A clear example was when the speakers were of opposite genders; the model gave better separation results at the perceptual and metrical levels.

The perceptual evaluation was carried out through a survey conducted with 23 listeners using four audio samples, two of which were from people with perceptually similar voices, while the others were from different voices. Each respondent was asked to answer ``How would they rate the quality of the speaker separation achieved by the model?''. The answers were given on a scale of 1 to 5, where 1 means no speaker separation and 5 means perfect speaker separation. Each sample contained the mixed audio, and the estimated speaker 1 and 2 separately.

In order to define which speakers are considered perceptually similar, the cosine similarity between speech embeddings from the speakers' utterances was estimated. Audios with perceptually similar voices obtained a cosine similarity between $[0.07 and 0.16]$, while the audios with perceptually different voices had a cosine similarity of $[0.03 and 0.07)$. All similarities were calculated using the Speech Embedding Wav2Vec \cite{baevski2020wav2vec} model described below.

A correlation analysis between the speaker similarity and the average rating of the separation quality shows a value of -0.7, a clear inverse relationship between speaker similarity and separation results. The higher the similarity, the lower the favorable results of a suitable separation at the perceptual and metric levels. 

Consequently, to improve the final speaker separation, we decided to include a term in the model's loss function, which counts on the speaker similarity. This idea follows a Domain-Adversarial Training approach \cite{ganin2016domain} where the gradient reversal layer is replaced by a speech embedding pre-trained architecture. 

Two different pre-trained Speech Embedding models were used: Wav2Vec and Pyannote; a quick review of them is presented in the following section. 

\subsection{Speech Embeddings}

Automatic Speech recognition (ASR) is a task that has made significant progress thanks to the development of deep learning. However, the training of speech recognition systems requires thousands of hours of transcribed speech to achieve acceptable performance \cite{baevski2020wav2vec}. A requirement that is simply difficult to have for the nearly 7,000 languages spoken in the world.

Due to this specific need, the Wa2Vec models and variants have been proposed: Wav2Vec 2.0 and Wav2Vec-U, which can be trained following a self-supervised approach for speech representation learning, to perform automatic speech recognition using unprocessed and unlabeled audios (without the transcription). Even though they were trained to accomplish an ASR task, in this work, we used them to estimate speaker similarity following previous approaches that have validated their use in speaker verification tasks \cite{fan2020exploring}. 

\subsubsection{Wav2Vec 2.0}

Wav2Vec 2.0 encodes the speech audio through a multilayer convolutional neural network, and then masks spans of the resulting latent speech representations, similar to masked language modeling. The latent representations are entered into a transformer network to build contextualized representations. The model is trained through a contrast task, in which it is necessary to distinguish between the true latent representation and the distractors. This model seems to solve the problem of the absence of labeled data since with only one hour of labeled data, Wav2Vec 2.0 outperforms a state-of-the-art ASR model trained with a subset of 100 hours \cite{baevski2020wav2vec}. Moreover, with only ten minutes of labeled data and a pre-training with 53,000 hours of unlabeled data, this model is able to achieve a word error rate (WER) of 4.8/8.2 \cite{baevski2020wav2vec}. 
Wav2Vec was trained using the Librispeech \cite{panayotov2015librispeech} corpus without transcriptions, which contains 960 hours of audio. It also used the LibriVox (LV-60k) corpus, which after applying some pre-processing operations \cite{kahn2020libri} was left with 53.2k hours of audio. 

During the experimentation process authors used 5 different data configurations: 960, 100, 10, 10, 1 and 0.1 hour of labeled data. This work used the model trained with the first configuration, which is freely available\footnote{\href{https://huggingface.co/docs/transformers/model_doc/wav2vec2}{wav2vec2}}.

%\footref{wav2vec2_link}. 

%\footnote{\href{https://huggingface.co/docs/transformers/model_doc/wav2vec2}{wav2vec2}\label{wav2vec2_link}}

\subsubsection{Pyannote}

Pyannote\cite{Bredin2020} is an open source toolkit written in Python for speaker diarization. Based on the Pytorch\texttrademark  framework, it contains pre-trained models covering a wide range of domains for speech activity detection, speaker change detection, voice overlap detection and voice embedding. The latter is the one we will use to generate the representative vectors of the speakers in order to compute a similarity between them.

Specifically, the embedding model available in Pyannote is based on the canonical Time Delay Neural Network (TDNN) architecture of \textbf{\textit{x-vectors}}, but with filter banks replaced by trainable SincNet functions \cite{ravanelli2018speaker}.

\subsection{Enhanced Conv-TasNet}\label{sec:enhanced}

\begin{figure}[!h]
    \centering
    \includegraphics[scale=0.6]{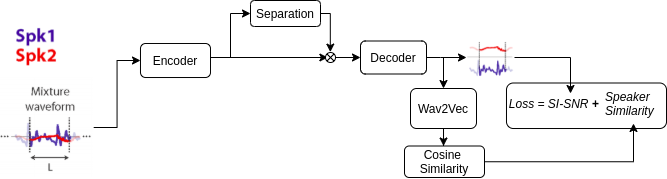}
    \caption{Key components of the modified Conv-TasNet architecture using Wav2Vec. Adapted from \cite{luo2019conv}.}
    \label{fig:convmodific}
\end{figure}

Figure \ref{fig:convmodific} shows a general diagram of the proposed modification of the Conv-TasNet architecture. The enhanced Conv-TasNet (E-Conv-TasNet)\footnote{\href{https://github.com/DW-Speech-Separation/train-test-ConvTasNet}{Enhanced Conv-TasNet GitHub repository}} adds a module that takes the decoder's output, separated Speakers 1 and 2, and calculates the embedding vectors (Wav2Vec or Pyannote) that, in turn, are used to estimate the cosine similarity. The cosine similarity term is added to the loss function similarly to a gradient reversal layer in a Domain-Adversarial training approach \cite{ganin2016domain}.

The hypothesis is that including such a term within the cost function could improve the performance of the original model since, during the training process, it would indicate to the model which speakers require more effort to be separated. 

%By adding such a similarity term to the base cost function, we would penalize those training samples, whose speakers have a higher similarity. The higher the similarity between speakers, the higher the error contribution.

Formally, given two speech embedding vectors ${\bf{x}}_1$ and ${\bf{x}}_2$ representing the speakers involved in the audio, the Cosine Speech Similarity (CSS) is defined as:

\begin{equation}
\text { CSS}=\frac{{\bf{x}}_1 \cdot {\bf{x}}_2}{\max \left(\left\|{\bf{x}}_1\right\|_{2} \cdot\left\|{\bf{x}}_2\right\|_{2}, \epsilon\right)}
\end{equation}

where $\epsilon$ is used to avoid division by zero.

Furthermore, the speaker similarity term to be included in the loss function follows a cross-entropy formulation of the form

\begin{equation}
    \text {Speech Similarity} = -WeightSL*\log_{} \left( \frac{1- \text {CSS}}{2} \right)
\end{equation}

where $WeightSL$ is a weight to balance the importance of the similarity component within the overall cost function. 

Finally, the complete expression defining the cost function is:

\begin{equation}
     \text {Loss ConvTasNet} =  \text {SI-SNR} +  \text {Speech Similarity}
\end{equation}

Where SI-SNR, is the error measure proposed in the original Conv-TasNet model.

\section{Experiments and results}\label{sec:exper}

\subsection{Corpora of phone calls}
In order to train these models, it is necessary to have a data set composed of 3 audios per sample: one audio containing the mix or recording of the call and two audios with the individual interventions of the speakers. It is essential to clarify that all audios must have the same duration.

After searching on the web, three sets of Spanish call recordings (corpora) were found: CallFriend, CallFriend-Caribbean-Accent-Spanish, and CallHome \cite{macwhinney2010transcribing}, which are provided by TalkBank, a project led by Brian MacWhinney at the Carnegie Mellon University, whose objective is to promote fundamental research in the study of human communication, with emphasis on spoken communication \cite{macwhinney2007talkbank}. These datasets contain call recordings in mp3 and wav formats, recorded at a sampling rate of 8KHz and in stereo channels (two channels), which allows obtaining the sources or individual interventions of the speakers easily. Table \ref{tbl:dataset} describes the duration of each of the data sets.

\begin{table}[]
\begin{center}
\caption{Corpus of speakers and duration of the datasets used for testing.}
\label{tbl:dataset}
\begin{tabular}{l|l|l}
\hline \hline
\textbf{id} & \textbf{Corpus name}              & \textbf{Time(Hours)} \\ \hline
1           & CallFriend-Spanish(CF)              & 36.6                     \\ \hline
2           & CallFriend-Caribbean-Accent-Spanish & 26.0                       \\ \hline
3           & CallHome-Spanish                    & 32.4                     \\ \hline
4           & CallFriend-Caribbean-CallHome (CFC-All) & 95.0                    \\ \hline \hline
\end{tabular}
\end{center}

\end{table}

The corpora CallFriend-Spanish and CallFriend-Caribbean-Accent-Spanish consist of 60 unscripted telephone conversations between native Spanish speakers. The recorded conversations last a maximum of 30 minutes. These datasets were obtained thanks to a campaign conducted through the Internet, publications (advertisements), and personal contacts. In this campaign, 100 people per dialect participated, who initiated the calls, each one made a single phone call, and most of the participants called family members or close friends. Upon successful completion of the call, the caller received "a payment of \$20. This dataset had two human audits; in the first one, it was verified that the correct language was being spoken and the quality of the recording. The second audit was performed by a native speaker familiar with the dialectal variation of Spanish to label the conversations as "Caribbean" or "non-Caribbean," according to the particular speech attributes of the participants.

On the other hand, the CallHome-Spanish corpus was obtained in the same way through a campaign; however, in this campaign, 200 callers participated. Each person was allowed to speak for up to 30 minutes and also received a payment of \$20 after making the call.

Finally, the dataset that we have named CallFriend-Caribbean-CallHome (CFC\_All) is the union of the three datasets. This dataset was created aiming to experiment with the behavior of the models as the amount of data increases. 

\begin{table}[]
\begin{center}
\caption{Corpora distribution.}
\label{tbl:dataset_division}
\begin{tabular}{l|l|l|l|l}
\hline \hline
\multicolumn{1}{c|}{\textbf{Corpus name}} & \multicolumn{1}{c|}{\begin{tabular}[c]{@{}c@{}}Train\\ (Hours)\end{tabular}} & \multicolumn{1}{c|}{\begin{tabular}[c]{@{}c@{}}Validation\\ (Hours)\end{tabular}} & \multicolumn{1}{c|}{\begin{tabular}[c]{@{}c@{}}Test\\ (Hours)\end{tabular}} & \multicolumn{1}{c}{\begin{tabular}[c]{@{}c@{}}Total time \\ (Hours)\end{tabular}} \\ \hline
CallFriend-Spanish(CF)                       & 26.40                                                                                 & 2.90                                                                               & 7.30                                                                           & 36.60                                                                                  \\ \hline
CallFriend-Caribbean-Accent-Spanish          & 18.20                                                                                 & 2.00                                                                                 & 5.80                                                                           & 26.00                                                                                    \\ \hline
CallHome-Spanish                             & 24.10                                                                                 & 2.42                                                                              & 5.91                                                                          & 32.40                                                                                  \\ \hline
CallFriend-Caribbean-CallHome (All)          & 68.68                                                                                & 7.32                                                                              & 19.01                                                                         & 95.01                                                                                 \\ \hline

\end{tabular}
\end{center}
\end{table}

\subsection{Experimental setup}

Each of the 4 data sets presented in the former section was divided into three different subsets: training, validation, and test, with a distribution of 70 \%, 20 \%, and 10 \%, respectively, of the total size of the dataset. Since the audios of the corpus are of long duration, they were split into 30-second portions.

\begin{figure}[!h]
    \centering
    \includegraphics[scale=0.58]{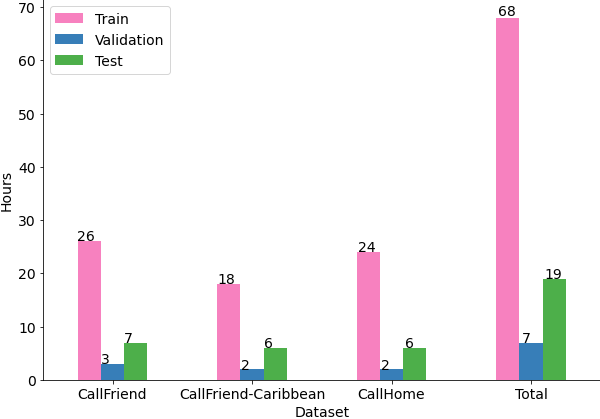}
    \caption{Distribution in hours of Spanish call recording corpus}
    \label{fig:distri_corpus}
\end{figure}

Table \ref{tbl:dataset_division} shows the distribution in hours of the subsets of data created for each of the corpora. As its name indicates, the training set is used to train the model, the validation set is used to select hyper-parameters, and finally, the test set allows us to know the model's performance with samples not seen by the model in any of the former phases. 

Figure \ref{fig:distri_corpus} shows the distribution of the four corpora in hours for each of the subsets (training, validation, and test). Finally, since several samples could correspond to segments from the same speakers, a Group Shuffle Split validation methodology was used to guarantee that all samples from the same speaker are only in one of the three subsets (training, validation, or test). Otherwise, the model could provide over-optimistic performance measures. 

\subsection{Results using original Conv-TasNet}

Initially, we trained the baseline model from scratch for 200 epochs, however, we had an early stop at epoch 146, obtaining a 9 dB performance metric, as shown in Table \ref{tbl:train_conv_CF}.

\begin{table}[]
\begin{center}
\caption{Conv-TasNet models trained using CallFriend-Spanish corpus for different number of epochs}
\label{tbl:train_conv_CF}
\begin{tabular}{c|c|c|c|c}
\hline \hline
\textbf{Training type} & \textbf{\begin{tabular}[c]{@{}c@{}}Training epochs\end{tabular}} & \textbf{\begin{tabular}[c]{@{}c@{}}Trained \\epochs\end{tabular}} & \textbf{SI-SDR} & \textbf{Best epoch} \\ \hline
From scratch                  & 200                                                                         & 146                                                                   & 9.035           & 116                  \\ \hline
Transfer-Learning           & 10                                                                          & 10                                                                    & 9.50            & 8                    \\ \hline
Transfer-Learning           & 20                                                                          & 20                                                                    & 11.51           & 17                   \\ \hline
Transfer-Learning           & 50                                                                          & 50                                                                    & 12.21           & 47                   \\ \hline
Transfer-Learning           & 100                                                                         & 100                                                                   & 13.18           & 90                   \\ \hline
Transfer-Learning           & 200                                                                         & 64                                                                    & 12.29           & 34                   \\ \hline \hline
\end{tabular}
\end{center}
\end{table}

Since it was not possible to obtain a significant improvement by training from scratch, we performed 5 training sessions starting from the wights of the available pre-trained model. 

\begin{figure}[!h]
    \centering
    \includegraphics[scale=0.7]{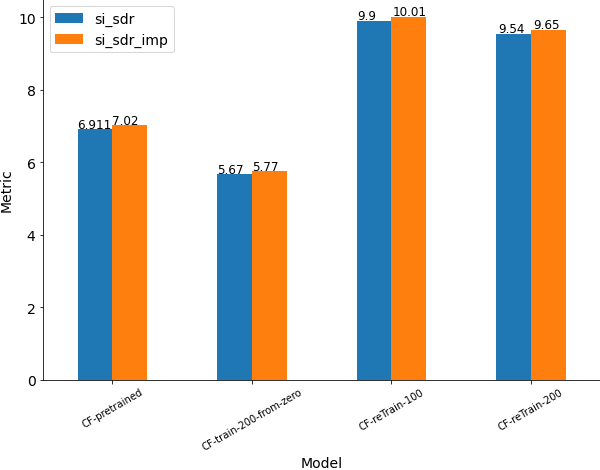}
    \caption{Conv-TasNet model performance trained with CallFriend-Spanish.}
    \label{fig:si-sdr-CF}
\end{figure}

As can be seen in Table \ref{fig:si-sdr-CF}, just by training the model for ten epochs using transfer learning, we managed to outperform the model trained from scratch by 146 epochs. This clearly shows the success of this approach. Notwithstanding the former results, more training experiments were performed increasing the number of epochs. The model trained for 100 epochs and using transfer learning is the one that obtains the best results, with an SI-SDR of 13.18 dB, followed by the model trained for 200 epochs, which obtains an SI-SDR value of 12.29 dB. It should be noted that these metrics were obtained using the validation set of the CallFriend-Spanish corpus.

After identifying the two best model configurations, we evaluated them using the test data set of the CallFriend-Spanish corpus, data that none of the models had seen before. With these results, we are able to determine if there is a real improvement in the models' performance compared to the one provided by the authors and which achieved an SI-SDR of 6.9 dB.

As can be seen in Figure  \ref{fig:si-sdr-CF}, the best model is the one trained using Transfer Learning for 100 epochs (CF-reTrain100). It obtained an SI-SDR of approximately 10 dB, exceeding the model provided by the authors (CF-pretrained) by 3 dB. 

%\subsection{Experiments using the CallFriend-Caribbean-CallHome (All) corpus}
After training with the CallFriend-Spanish corpus, we decided to perform two experiments using the CFC\_ALL corpus. Initially, the model was trained from scratch for 200 epochs, however, the training had an early stop at epoch 102, since no improvement in the validation metric was achieved, obtaining an SI-SDR of 13.62 dB as shown in Table \ref{tbl:train_all}. Then, transfer learning was used again, training the model for 200 epochs. There was an early stopping during training at epoch 93, achieving an SI-SDR of 16.13 dB in the validation set, which is approximately 3 dB better than training the model from scratch. 

\begin{table}[]
\begin{center}
\caption{Performance of the Conv-TasNet model training using CFC\_All corpus.}
\label{tbl:train_all}
\begin{tabular}{|c|c|c|c|c|}
\hline
\textbf{Training type} & \textbf{\begin{tabular}[c]{@{}c@{}}Training \\ epochs\end{tabular}} & \textbf{\begin{tabular}[c]{@{}c@{}}Trained \\ epochs\end{tabular}} & \textbf{SI-SDR} & \textbf{Best epoch} \\ \hline
From scratch                  & 200                                                                         & 102                                                                   & 13.62           & 72                   \\ \hline
Transfer-Learning           & 200                                                                         & 93                                                                    & 16.13           & 63                   \\ \hline
\end{tabular}
\end{center}
\end{table}

\begin{figure}[!h]
    \centering
    \includegraphics[scale=0.7]{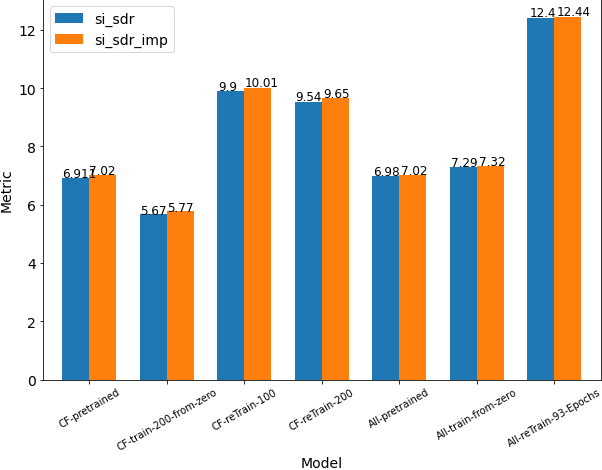}
    \caption{Performance of Conv-TasNet trained with CallFriend-Caribbean-CallHome (CFC\_All) corpus.}
    \label{fig:si-sdr-ALL}
\end{figure}

Figure \ref{fig:si-sdr-ALL} shows the results of using the CFC\_ALL test set to evaluate the two models, the one trained from scratch and the model trained using transfer learning. The first model obtained an SI-SDR of 7.29 dB, while the second got an SI-SDR of 12.4 dB. In addition, the pre-trained model without any modification obtained an SI-SDR of 6.98 dB. 

Although the models trained with the CFC\_ALL and CallFriend-Spanish corpus cannot be compared with each other since they were trained and validated with different information sources, by analyzing some audios estimated by them we found that the audios generated by the model trained with the CFC\_ALL corpus include noise artifacts, therefore the model trained with Transfer Learning and the CallFriend-Spanish corpus (CF-reTrain-100) showed a better performance according to perceptual criteria.

\begin{figure}[htbp]
\centering
\subfloat[]{\includegraphics[width=7.5cm]{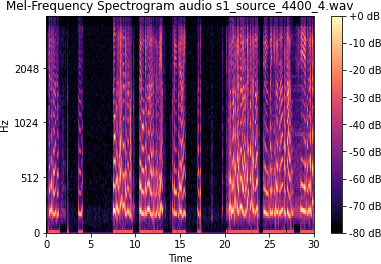}
}
\subfloat[]{\includegraphics[width=7.5cm]{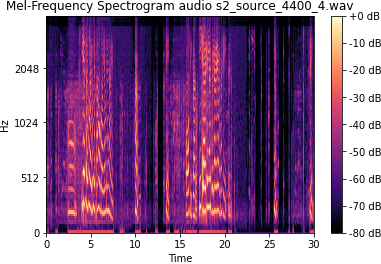}
}

\hspace{0mm}
\subfloat[]{\includegraphics[width=7.5cm]{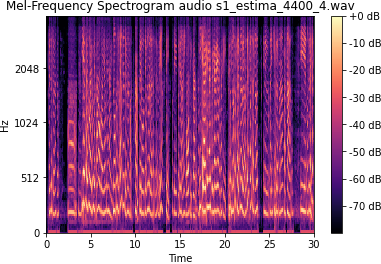}
}
\subfloat[]{\includegraphics[width=7.5cm]{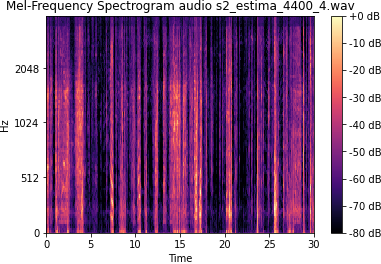}
}
\caption{ Comparison of estimated and original spectrograms of an example audio with poor separation.}
\label{fig:especto_bad}
\end{figure}

Regarding the analysis of the models' performance when the speakers involved have perceptually similar voices, figure \ref{fig:especto_bad} shows the Mel-scale spectrograms of the clean and estimated source of an audio call involving two women in this condition. Graphically, we can see that the estimates are far away from the clean sources. The model obtained a performance of -9.14 dB for the SI-SDR metric, which is unfavorable.

\begin{figure}[htbp]
\centering
\subfloat[]{\includegraphics[width=7.5cm]{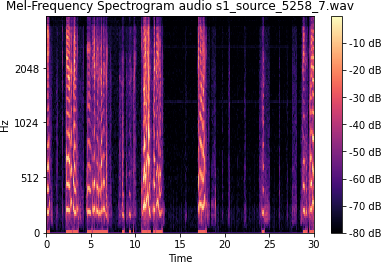}
}
\subfloat[]{
\includegraphics[width=7.5cm]{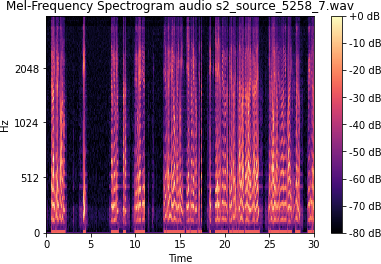}
}
\hspace{0mm}
\subfloat[]{
\includegraphics[width=7.5cm]{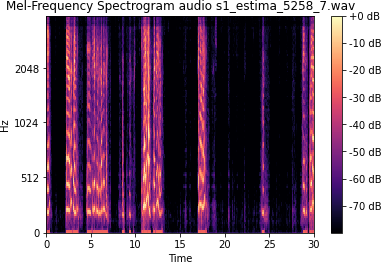}
}
\subfloat[]{
\includegraphics[width=7.5cm]{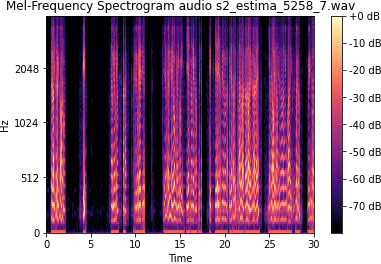}
}
\caption{ Comparison of estimated and original source spectrograms of an example audio with good separation.}
\label{fig:especto_good}

\end{figure}

On the other side, when the voices are perceptually different, the results are favorable, a case that can be seen clearly in Figures \ref{fig:especto_good}, where the estimated and clean spectrograms are graphically similar. In addition, at the level of the separation metric (SI-SDR), a value of 21.74 dB was obtained in this particular case.

Based on the previous observations, it was decided to make a modification to the original architecture, thus creating the enhanced ConvTasNet described in section \ref{sec:enhanced} and whose results are shown below.

\subsection{Results using E-Conv-TasNet model}

The proposed modification to the Conv-TasNet model requires an experimental phase in which different values of $WeightSL$ and different representative vectors were evaluated. This is because the Wav2Vec 2.0 model has 12 different outputs corresponding to intermediate layers of the model, so there are 12 different representations of the same audio. Additionally, Pyannote embedding vectors were also explored. 

A total of 15 experiments were performed using Wav2Vec as embedding, where E-Conv-TasNet models were trained using transfer learning. All models were trained for 100 epochs. Each model has a different $WeightSL$ value in the grid [5, 10, 20]. Moreover, since the Wav2Vec model provides 12 possible representative vectors, outputs of the 1, 2, 3, 4, and 12 layers were evaluated. By combining the possible values of these two hyper-parameters, we end up with a total of 15 experiments. Finally, all the experiments were performed using the CallFriend-Spanish corpus.

The initial objective of these experiments was to determine whether there is a real improvement regarding separation and perceptual metrics by introducing the speaker similarity term. In addition to this, the optimal configuration for $WeightSL$ and the Speech Embedding layer was also analyzed.

\begin{figure}[!h]
    \centering
    \includegraphics[scale=0.5]{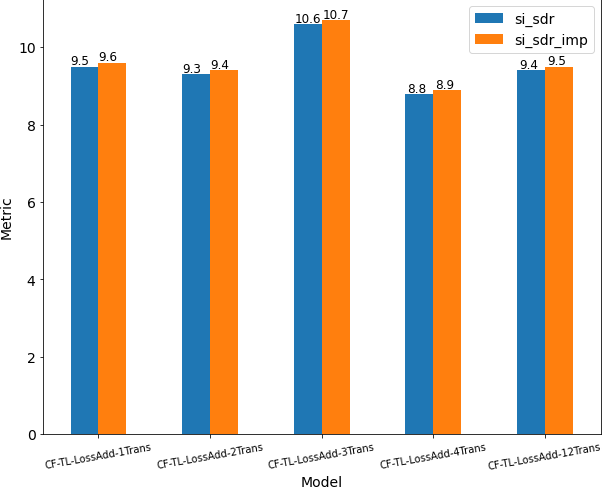}
    \caption{Results of the  E-Conv-TasNet model trained with CallFriendSpanish and a $WeightSL$ of 5.}
    \label{fig:si-sdr-add_loss_5}
\end{figure}

In the first five experiments, the $WeightSL$ value was set to 5, and the number of the Speech Embedding layer was permuted within the possible values (1,2,3,4,12). As shown in Figure \ref{fig:si-sdr-add_loss_5}, the best result obtained (10.6 dB in SI-SDR) was achieved using embeddings from layer 3. We will name this model as CF-TL-LossAdd-3-Trans-5-Weight. This result outperforms the original CF-reTrain-100 model, which obtained an SI-SDR of 9.9. 

\begin{figure}[!h]
    \centering
    \includegraphics[scale=0.48]{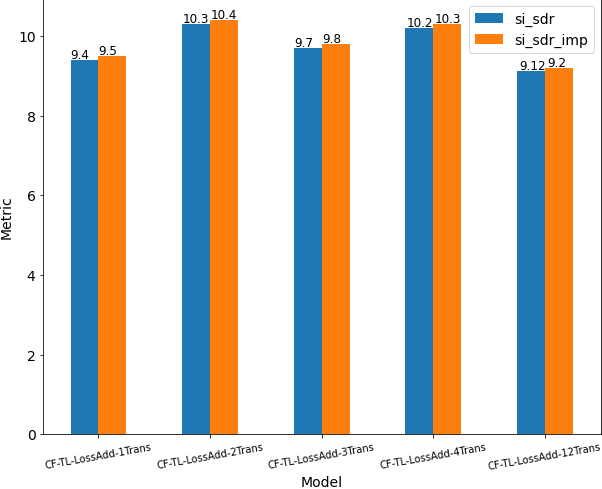}
    \caption{Results of the E-Conv-TasNet model trained with CallFriendSpanish and a $WeightSL$ of 10. }
    \label{fig:si-sdr-add_loss_10}
\end{figure}

\begin{figure}[!h]
    \centering
    \includegraphics[scale=0.49]{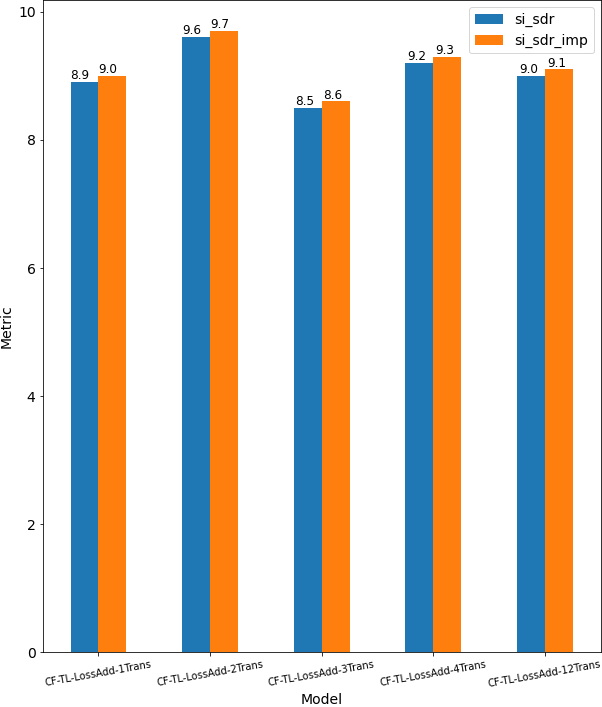}
    \caption{Results of the E-Conv-TasNet model trained with CallFriendSpanish and $WeightSL$ of 20.}
    \label{fig:si-sdr-add_loss_20}
\end{figure}

In the subsequent five experiments, the value of $WeightSL$ was set to 10, obtaining two models with SI-SDR above 10 dB; these models were those trained with Speech Embedding layers 2 and 4 and obtained a metric of 10.3 and 10.2 dB, respectively (see Figure \ref{fig:si-sdr-add_loss_10}). For the last five experiments, the $WeightSL$ value was set to 20, but none of the models achieved an SI-SDR value higher than 10 dB, as shown in Figure \ref{fig:si-sdr-add_loss_20}.

Finally, we replicated the experiments using Pyannote as speech embedding and set the $WeightSL$ to 5, obtaining an SI-SDR of 9.06 dB. Using a $WeightSL$ of 10, we obtained an SI-SDR of 8.7 dB. These results show a superior performance of Wac2Vec for the task and settings evaluated in this work.

\subsection{Deployment}

In order to preliminarily evaluate the limitations of the model in a real-time environment, a basic prototype based on the sound device library \cite{sounddevice} was developed. 

%In order to evaluate preliminarly the limitations of the model in a real time environment, a basic prototype was developed. For the deployment of the model in real time, the sounddevice library was used; it provides us with an API to play and record Numpy matrices containing audio signals. Available for operating systems such as: Linux, macOs and Windows.

%Moreover, the Tkinter library was used to create a user-friendly graphical interface and to be able to use the model in real time. As shown in Figure \ref{fig:gui}, the graphical interface has two simple buttons: a button to start recording, which when is pressed it  will start capturing the audio from the microphone and will internally run the model in real time, reproducing at the same time the interventions of each speaker through the audio output, using a specific channel for each speaker (left or right earphone).

%\begin{figure}[!h]
%    \centering
%    \includegraphics[scale=0.5]{Figures/arquitecturas/GUI.png}
%    \caption{Graphical user interface of the model displayed in real time.}
%    \label{fig:gui}
%\end{figure}

\begin{figure}[!h]
    \centering
    \includegraphics[scale=0.38]{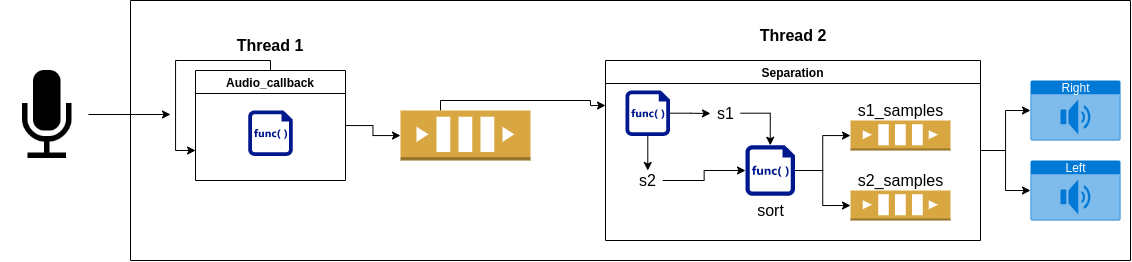}
    \caption{Diagram showing how the model works in real time.}
    \label{fig:deploy}
\end{figure}

Figure \ref{fig:deploy}, illustrates a general diagram of the prototype's information flows. It starts with the real-time audio capture, which is handled by a function (Audio callback) running in Thread 1; this function is responsible for receiving the different segments of the audio in real-time, storing them in a queue, and then be processed by the Separation function one by one in order of arrival.

The Separation function in Figure \ref{fig:deploy} has two sub-processes: the first one is in charge of separating the interventions of the speakers of the segment at hand, and the second one takes the two generated audios and assigns them to their corresponding location (vector of interventions speaker one or vector of interventions speaker two), depending on the similarity that exists between these segments and particular speakers segments used as reference. The reference segments have a duration of 1 second and are recorded at the beginning of the process. Finally, the speakers' audio is played through different channels and preserving the order; that is, if speaker one is played through channel 1 (left earphone), all the interventions of the same speaker should be heard through the same channel.

Once the architecture mentioned in Figure \ref{fig:deploy} was set up, we performed an experiment to measure the performance of the system using 40 samples of the test set of the CallFriend-Spanish(CF) corpus. The separation of speakers was performed by changing the size of the segment to be processed from 1 to 10 seconds.

\begin{figure}[!h]
    \centering
    \includegraphics[scale=0.7]{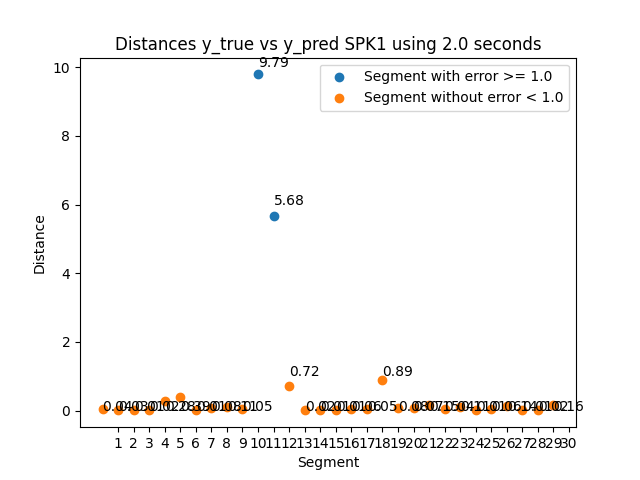}
    \caption{Distances between 2-second-long segments in one speaker.}
    \label{fig:dist_s1}
\end{figure}

In order to estimate the number of times the prototype fails to identify the correct channel for each speaker, the Euclidean distance between the clean and the estimated audio segments was calculated. If this distance exceeds a previously defined threshold (1.0), the estimated segment was misplaced, i.e., there is a synchronization error. The value of this threshold was experimentally determined as follows:

\begin{enumerate}
\item Forty random samples were selected from the CallFriend-Spanish corpus test set.
\item Each sample was framed using different length segments (1,2,...,10, 15, 20, 25 and 30 seconds).
\item The distances between the clean and estimated speech signals for each speaker were plotted, as shown in figure \ref{fig:dist_s1}.
%\item A visual comparison of the clean and estimated waveform was made to determine if the estimated segments above this threshold were different from the true ones.
\item Using the graph \ref{fig:dist_s1}, a threshold of 1.0 was found as the minimal difference to identify a misplaced speech segment.
\end{enumerate}

A concrete example of how the segment synchronization error is calculated is the following. Figure \ref{fig:dist_s1} show the clean and estimated waveform of two speakers present in a call, it is possible to visually appreciate a high degree of similarity between both, however there are two segments whose distance exceeds the threshold (9.79 and 5.68). We can see in figure \ref{fig:dist_s1}, that those segments contains a synchronization error. The corresponding synchronization error is 6.6 \% (2/30 segments).

\begin{figure}[!h]
    \centering
    \includegraphics[scale=0.7]{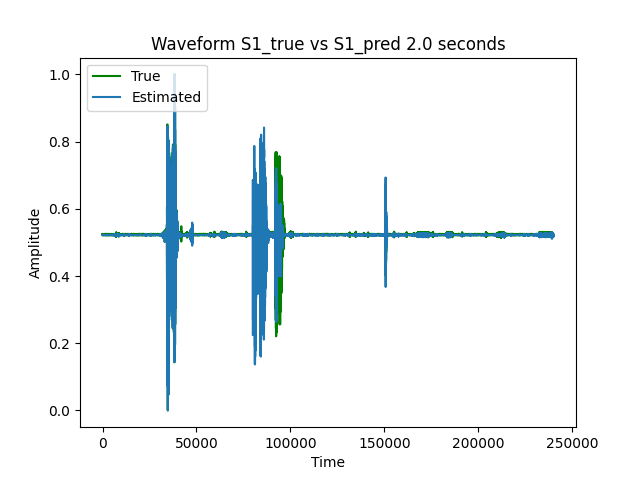}
    \caption{Estimated waveform vs. clean waveform using 2 second long segments, speaker number 1.}
    \label{fig:s1}
\end{figure}

\begin{figure}[!h]
    \centering
    \includegraphics[scale=0.7]{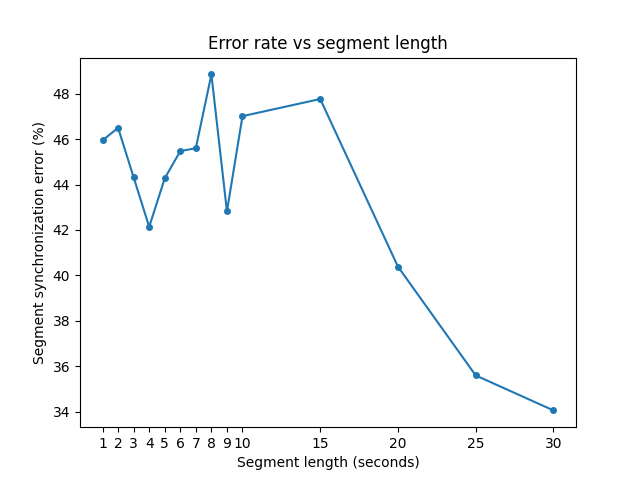}
    \caption{Average synchronization error vs. segment length.}
    \label{fig:error_longitude}
\end{figure}

Figure \ref{fig:error_longitude} shows a clear trend of an inverse relationship between the length of the segment to be processed and the percentage of error. This quantitatively evidences the low performance of the system using segments of 1-second length. A longer segment length will allow us to have better results but mines the aim of real-time processing.

\section{Conclusions}\label{sec:coclusions}

The use of Transfer Learning during the training process of the Conv-TasNet model significantly improved the speaker separation results, going from a metric (SI-SDR) of 6.9 dB to 9.9 dB, improving the quality of the separation by 3 dB. This result reaffirms the great usefulness of this technique during the training process, even when the pre-training process was carried out in a different language.

The Enhanced Conv-TasNet model obtained the best performance (10.6 dB in SI-SDR) during this work's experiments. The modification to the cost function using the inter-speaker similarity term, based on the cosine similarity between the representative vectors given by the Speech Embedding Wav2Vec model, improved the model's performance by 0.7 dB. This result was achieved using layer 3 of the Wav2Vec model, which is also a model trained for the English language. Additional experiments should be carried out with a large Spanish corpus. 

All models were trained, adjusted, and validated with audios of 30 seconds. However, when testing the model in real-time conditions, we used segments of 1 second, which explains its limited performance.

The architecture evaluated for the deployment allowed the consumption of the model in real-time; however, given the short duration of the processed segments, for several segments, only one speaker was present in the recording, turning the task into a speaker identification rather than a speech separation problem. This fact also brings an additional challenge since the assignment of the speaker channel is an unsupervised task, and the system must guarantee that every segment of a speaker is placed in the right channel. In this respect,  the results showed that there is an inverse relationship between the length of the segment to be processed and the percentage of synchronization error. Alternative solutions to this problem should be addressed in future works.

\section*{Acknowledgment}
J. Arias-Londoño started this work at the Antioquia University and finished it supported by a María Zambrano grant from the Universidad Politécnica de Madrid, Spain.

%Bibliography
\bibliographystyle{unsrt}

\end{document}